\documentclass[
twocolumn,%
nofootinbib,%
preprintnumbers,%
showpacs,%
prd]{revtex4}
\usepackage{graphicx}
\newcommand{\be}{\begin{equation}}
\newcommand{\ee}{\end{equation}}
\newcommand{\bea}{\begin{eqnarray}}
\newcommand{\eea}{\end{eqnarray}}

\newcommand{\geff}{g_{\rm eff}}

\newcommand{\muMS}{\bar\mu_{{\rm MS}}}

\newcommand{\rmi}[1]{{\mbox{\scriptsize #1}}}

\begin{document}
\preprint{TUW-03-35,
HIP-2003-56/TH}
\title{Perturbative QCD 
at non-zero
chemical potential:
\\
Comparison with the 
large-$N_f$ limit and apparent convergence
}
\author{A. Ipp}
\affiliation{Institut f\"ur Theoretische Physik, Technische
Universit\"at Wien, \\Wiedner Haupstr.~8-10, 
A-1040 Vienna, Austria }
\author{A. Rebhan}
\affiliation{Institut f\"ur Theoretische Physik, Technische
Universit\"at Wien, \\Wiedner Haupstr.~8-10, 
A-1040 Vienna, Austria }
\author{A. Vuorinen}
\affiliation{Department of Physical Sciences and
Helsinki Institute of Physics \\
P.O. Box 64, FIN-00014 University of Helsinki, Finland}
\begin{abstract}
The perturbative three-loop result for the thermodynamic potential
of QCD at finite temperature and chemical potential as obtained
in the framework of dimensional reduction is compared
with the exact result in the limit of large flavor number.
The apparent convergence of the former as well as possibilities for
optimization are investigated. Corresponding optimized results
for full QCD are given for the case of two massless quark flavors.  
\end{abstract}
\maketitle

\section{Introduction}

Quite some progress has been made recently in extending
perturbative and nonperturbative results on the thermodynamic
potential of hot deconfined QCD to nonzero chemical potential.
On the lattice, where a nonzero quark chemical potential
gives rise to a computationally problematic complex action,
a number of new techniques have been devised that
make it possible to study the effects of a not too large quark
chemical potential $\mu\lesssim T$,
and first results have been produced for the equation of state
\cite{Fodor:2002km,Gavai:2003mf,Allton:2003vx}.
Concurrently, the perturbative series expansion 
of the QCD pressure, which 
had
been determined up to and including
the last fully perturbative order $g^6\ln g$
in
Refs.~\cite{Arnold:1995eb,Zhai:1995ac,Braaten:1996jr,Kajantie:2002wa},
has been extended to non-zero chemical potential in
Refs.~\cite{Hart:2000ha,Vuorinen:2002ue,Vuorinen:2003fs}
within the framework of dimensional reduction
\cite{Braaten:1996jr,Kajantie:1996dw}. 
Dimensional reduction can be expected to work as long as $2\pi T$ is the
dominant energy scale. In the limit of large $\mu/T$ this
ceases to be the case 
when the scale $g\mu$ appearing in the
three-dimensional mass parameter $m_E$ becomes of the same order.
In the regime $T\lesssim g\mu$, perturbation theory requires
a reorganization with new phenomena such as the onset of
non-Fermi-liquid behavior \cite{Ipp:2003cj}, and for the parametrically
smaller temperatures $T\lesssim \mu g^{-5}\exp[-3\pi^2/(\sqrt2 g)]$
eventually the nonperturbative phenomenon of color superconductivity
arises \cite{Son:1998uk}.


Since at any temperature of practical or even theoretical relevance,
the QCD coupling $g$ is not much smaller than 1,
the predictivity of the perturbative results remains uncertain.
In fact, it is a well known problem of perturbative results
at finite temperature that the strictly perturbative series expansion
has extremely poor apparent convergence and large 
renormalization scale dependence up to temperatures of at least
$10^5$ times the deconfinement temperature $T_c$
\cite{Zhai:1995ac,Braaten:1996jr}.
On the other hand, it has been found that
already resummations involving only the lowest order
contributions 
from the soft scale $gT$ 
can improve the situation decisively (\cite{Blaizot:2003tw}
and references therein).
Within the framework of dimensional reduction, a significant improvement
can be achieved by the simple resummation provided by
keeping 
effective-field-theory
parameters unexpanded \cite{Kajantie:2002wa,Blaizot:2003iq}.

In particular at non-zero quark chemical potential, where lattice
data are still somewhat preliminary,
it would be helpful to be able to test the 
predictivity of the (resummed) perturbative results by other means.
Most recently the 
limit of large flavor number $N_f\to\infty$,
with finite $\geff^2\equiv g^2N_f/2$, 
has been developed as such a theoretical test bed.
In Refs.~\cite{Moore:2002md
}
the large-$N_f$ pressure has
been determined exactly
for all temperatures sufficiently below the scale of the Landau pole 
$\Lambda_L\sim T\exp(6\pi^2/\geff^2)$
that is present in the large-$N_f$ theory; in Ref.~\cite{Ipp:2003jy}
this has now been extended to non-zero chemical potential.
In this Brief Report, we use the results  of Ref.~\cite{Ipp:2003jy}
obtained
in the large-$N_f$ limit
to investigate the apparent convergence
of the perturbative results of Ref.~\cite{Vuorinen:2003fs}
obtained in dimensional reduction at non-zero
chemical potential. We are also able to study for the first
time quantitatively the inevitable breakdown of dimensional
reduction at large $\mu/T$ in the context of the pressure.
For moderate $\mu/T$, we consider the possibilities for optimization of
the perturbative results,
and present correspondingly
optimized results for $N_f=2$ massless flavors in full QCD.

\section{Large $N_f$}



In the limit of a large number of quark flavors $N_f$ 
with a common chemical potential $\mu$,
the dimensional reduction 
result for the interaction part of the pressure of QCD
at finite temperature and chemical potential obtained
in \cite{Vuorinen:2003fs} 
reduces to ($\geff^2\equiv g^2N_f/2$)
\begin{eqnarray}
\label{PLNf}
&&\left. \frac{P-P_{0}}{N_{g}}\right|^{\rm {DR}}_{N_f\to\infty} 
 =  -\left( \frac{5}{9}T^{4}+\frac{2}{\pi ^{2}}\mu ^{2}T^{2}+\frac{\mu ^{4}}{\pi ^{4}}\right) \frac{\geff ^{2}}{32}\nonumber \\
 &&\quad+\frac{1}{12\pi }m_{E}^{3}T
+\bar{\alpha }_{\rmi {E3}}\frac{\geff ^{4}}{(48\pi )^{2}}+O(\geff ^{6}),
\end{eqnarray}
where $N_g$ is the number of gluons.
The effective field theory parameter \( m_{E}^{2} \) is given
by
\bea
\label{mE2}
m_{E}^{2}&=&\left( T^2+\frac{3\mu ^{2}}{\pi ^{2}}\right) \biggl[ \frac{\geff ^{2}}{3}\nonumber\\&-&\!\!
\frac{\geff ^{4}}{(6\pi )^{2}}\left( 2\ln \frac{\muMS }{4\pi T}-1-\aleph (z)\right) \biggr] +O(\geff ^{6}),
\eea
and the coefficient of the
order $\geff^{4}$-term in (\ref{PLNf}) by
\begin{eqnarray}
\bar{\alpha }_{\rmi {E3}} & = & 12\left\{ \frac{5}{9}T^{4}+\frac{2}{\pi ^{2}}\mu ^{2}T^{2}+\frac{\mu ^{4}}{\pi ^{4}}\right\} \ln \frac{\muMS }{4\pi T}\nonumber \\
 &+& 4T^{4}\left\{ \frac{1}{12}+\gamma -\frac{16}{15}\frac{\zeta '(-3)}{\zeta (-3)}-\frac{8}{3}\frac{\zeta '(-1)}{\zeta (-1)}\right\} \nonumber \\
 &+& \frac{\mu ^{2}T^{2}}{\pi ^{2}}\left\{ 14+24\gamma -32\frac{\zeta '(-1)}{\zeta (-1)}\right\} +\frac{\mu ^{4}}{\pi ^{4}}(43+36\gamma )\nonumber \\
 &-&96T^{4}\left\{ 3\aleph (3,1)+8\aleph (3,z)+3\aleph (3,2z)-2\aleph (1,z)\right\} \nonumber \\
 &+&\frac{48i\mu T^{3}}{\pi }\left\{ \aleph (0,z)-12\aleph (2,z)-12\aleph (2,2z)\right\} \nonumber \\
 &+&\frac{96\mu ^{2}T^{2}}{\pi ^{2}}\left\{ 4\aleph (1,z)+3\aleph (1,2z)\right\} \nonumber \\
&+&\frac{144i\mu ^{3}T}{\pi ^{3}}\aleph (0,z)
\label{coefficient_aE3_largeNf} 
\end{eqnarray}
with $z\equiv \frac{1}{2}-i\frac{\mu }{2\pi T}$
and the special functions \cite{Vuorinen:2003fs}
\begin{eqnarray}
\aleph (n,w) & \equiv  & \zeta' (-n,w)+(-1)^{n+1}\zeta '(-n,w^{*}),\\
\aleph (w) & \equiv  & \Psi (w)+\Psi (w^{*}).
\end{eqnarray}
\( \zeta  \) is the 
Riemann zeta function,
$\zeta '(x,y)  \equiv   \partial _{x}\zeta (x,y)$,
and \( \Psi  \) the digamma function 
\( \Psi (w)\equiv \Gamma '(w)/\Gamma (w) \).
Note that despite the appearance of complex quantities in (\ref{coefficient_aE3_largeNf})
the coefficient \( \bar{\alpha }_{\rmi {E3}} \) is real as it has
to be.

To the order of accuracy of the result (\ref{PLNf}),
the relevant dimensionally reduced effective theory in the large-$N_f$ limit
turns out to be noninteracting, 
contributing only the term
$m_E^3 T$ in (\ref{PLNf}). This gives rise to nonanalytic terms
in $\geff^2$, namely single powers of $\geff$, but no logarithmic
terms as in full QCD.

\begin{figure}[t]
\vspace{2mm}
\centerline{\includegraphics[width=0.8%
\linewidth]{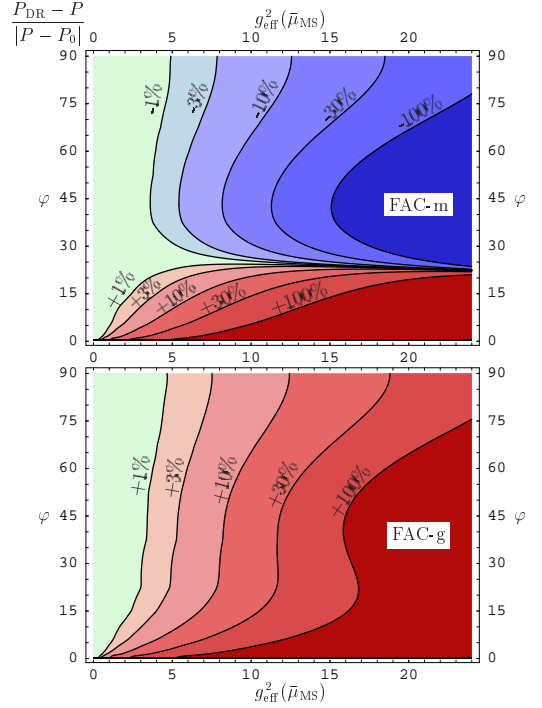}}
\caption{Percentage errors of the perturbative result for the
interaction part of the pressure to
order $\geff^5$ in the large-$N_f$ limit
for two choices of $\muMS$: Fastest apparent convergence
of $P$ as well as
$m_E^2$ (FAC-m), and of $g_E^2$ (FAC-g).
The brightest area corresponds to an error of less than 1\%,
the darkest ones to an error of over 100\%.
The ratio of chemical potential to temperature increases from
top to bottom according to $\varphi=\arctan(\pi T/\mu)$ so that
$90^\circ$ corresponds to high temperature and zero chemical
potential, and $0^\circ$ to zero temperature and high chemical potential.
The coupling is given in terms of
$\geff^2(\muMS)$ at $\muMS=\sqrt{\pi^2 T^2+\mu^2}$.
\label{fig:contpl}}
\end{figure}

Just as in full QCD, however,
the perturbative result has
unusually large renormalization scale dependence
for 
$\geff\gtrsim2$.
But, as noted in
Ref.~\cite{Moore:2002md
}, 
choosing the
renormalization scale $\muMS$ such that the $\geff^4$ correction
in $m_E^2$ as given in Eq.~(\ref{mE2}) is put to zero (FAC-m)
gives good agreement with the exact result
up to $\geff^2\sim 7$
at zero chemical potential. At large $N_f$,
this prescription is also FAC with respect to the pressure itself, since
it puts to zero the coefficient of its $\geff^5$-term.

In the upper panel of
Fig.~\ref{fig:contpl} the quality of the perturbative result
to order $\geff^5$ in the large-$N_f$ limit with $\muMS=\muMS^{\rm FAC-m}$
is displayed for the entire $\mu$-$T$ plane.
The deviation of the thus optimized perturbative result 
for the interaction part of the pressure 
from the exact one is shown in the form
of a contour plot 
of $[(P_{\rm DR}-P_0)-(P-P_0)]/|P-P_0|=(P_{\rm DR}-P)/|P-P_0|$,
where $P$ is the exact result from Ref.~\cite{Ipp:2003jy}
and $P_0$ the ideal-gas value.
The $\mu$-$T$ plane is
parametrized by $\varphi=\arctan(\pi T/\mu)$ and
$\geff^2(\muMS)$ at $\muMS=\sqrt{\pi^2 T^2+\mu^2}$. The accuracy of the
FAC-m result at zero chemical potential ($\varphi=90^\circ$)
is seen to decrease slowly with increasing chemical potential,
apart from an accidental zero of the error when
$\varphi \sim 22^\circ$; for slightly smaller
$\varphi\lesssim 18^\circ$, i.e.\
$T\lesssim 0.1\mu$, the errors eventually start to grow rapidly,
marking the breakdown of dimensional reduction.

While FAC-m is particularly natural in the large-$N_f$ limit,
one can consider different optimizations. At subleading orders
in the large-$N_f$ expansion, one would also need the effective-field-theory
parameter
\begin{equation}
g^{2}_{E}=T\left[ \geff ^{2}+\frac{\geff ^{4}}{3(2\pi )^{2}}\left( 2\ln \frac{\muMS }{4\pi T}-\aleph (z)\right) \right] +O(\geff ^{6})
\end{equation}
and one could choose to set 
$\muMS$ such that the $\geff^4$ correction
in $g_E^2$ 
is put to zero (FAC-g).
Curiously enough, the corresponding $\muMS$ 
happens to be strictly proportional to that of the FAC-m scheme,
$\muMS^{\mbox{\scriptsize FAC-g}}(T,\mu)=e^{-1/2}\muMS^{\mbox{\scriptsize FAC-m}}(T,\mu)$. 
In contrast to FAC-m, this does not minimize the $\geff^5$ coefficient
in the pressure. Nevertheless, as can be seen in the lower
panel of Fig.~\ref{fig:contpl},
the quality of the perturbative
result is rather similar to FAC-m: the sign of the deviation
is now positive throughout the $\mu$-$T$ plane so that there is
no accidental zero of the error, but
the absolute magnitude of the error is 
otherwise comparable.


Because FAC-m and FAC-g have comparable errors with opposite
sign for $\varphi\gtrsim 18^\circ$, the optimal choice
of renormalization
scale for the 3-loop result happens to lie in between the two FAC scales.
Incidentally, their arithmetic mean turns out
to give a surprising accuracy with errors below 1\% 
for all values of $\geff^2$ considered, if
$\varphi\gtrsim 67^\circ$ or $\mu/T\lesssim 1.3$.
However, for larger $\mu/T$ the range of coupling with
small errors shrinks quickly; for $T\lesssim 0.1\mu$
the errors are finally no smaller than
with either FAC-m or FAC-g.

Thus, except at rather small $T/\mu$, the two 
FAC optimizations
that we have considered
appear to give very satisfactory results
in the large-$N_f$ limit, even
at coupling $\geff^2$ so large that the renormalization scale
dependence is already quite strong.
The renormalization scales they lead to are in fact rather
close to each other---they differ just by a factor
of $e^{1/2}\approx 1.6$, and 
they (perhaps fortuitously) turn out to
enclose the actual optimal choice for $T/\mu \gg 0.1$.

\section{Finite $N_f$
}

It is of course far from guaranteed that
the optimizations FAC-m and FAC-g
will work equally well in full, i.e. finite-$N_f$
QCD (in the deconfinement phase). When applying these prescriptions now to the
case of $N_f=2$,
we shall therefore also consider the size of
the renormalization scale dependence as a further criterion
and investigate whether it can be reduced by a simple resummation
consisting of keeping effective-field-theory parameters
unexpanded.

In Ref.~\cite{Blaizot:2003iq}, it has been found that at zero chemical
potential such a resummation
considerably reduces the scale dependence. Using
a standard 2-loop running coupling one can even fix the renormalization
scale by a principle of minimal sensitivity (PMS), i.e.\
by requiring that the derivative of $P$ with respect to $\muMS$
vanishes, and this
turns out to be close to the FAC result.\footnote{In the large-$N_f$
limit PMS cannot be applied because there the scale dependence
turns out to be monotonic.}
While the formal
$\muMS$-independence of the perturbative
result to order $g^5$ or $g^6\ln g$ requires only a one-loop
beta function, it is not inconsistent to use a more accurate
coupling. We in particular adopt the 2-loop running coupling
because the QCD scale $\Lambda_{\rm QCD}$ of other renormalization
schemes is frequently related to the one of the $\overline{\hbox{MS}}$ scheme
using the 2-loop beta function. Also, the 2-loop coupling is
already reasonably close to the 3- or 4-loop coupling.

\begin{figure}[t]
\centerline{\includegraphics[viewport=70 230 540 810,width=0.79%
\linewidth]{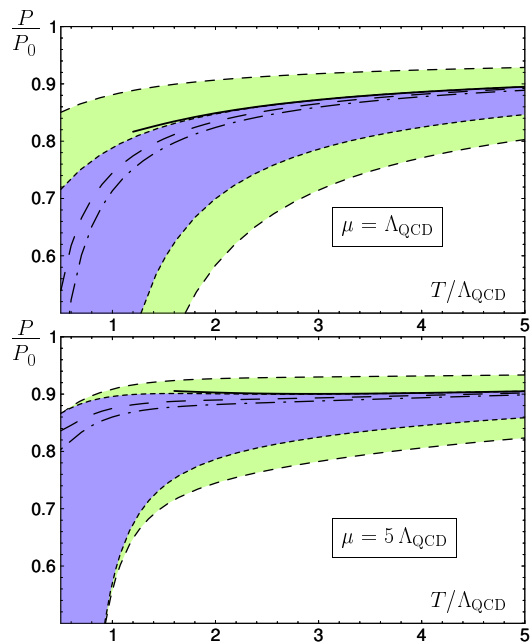}}
\caption{Three-loop pressure in deconfined QCD with two massless flavors
for $\mu=\Lambda_{\rm QCD}$ (upper panel) and $\mu=5\Lambda_{\rm QCD}$
(lower panel).
The bright-colored bands show the strictly perturbative results to
order $g^5$ with $\muMS$ varied about
the FAC-m value (dash-dotted line) by a factor of 2, the darker bands
the same when $m_E^2$ is left unexpanded.
The dashed line is the FAC-g result,
the full line beginning at $T\sim 1.5\Lambda_{\rm QCD}$ the PMS result.
\label{fig:res15}}
\end{figure}

In Fig.~\ref{fig:res15} we have repeated the analysis of
the scale dependence of \cite{Blaizot:2003iq} for the perturbative 
$N_f=2$ result of Ref.~\cite{Vuorinen:2003fs} to order\footnote{We
refrain, however, from extending this analysis to the order $g^6\ln g$
contribution as done in Ref.~\cite{Blaizot:2003iq} for
zero chemical potential, because it would involve an unknown function of
$\mu/T$ rather than a mere constant.}
$g^5$
with a quark chemical potential
of $\mu=\Lambda_{\rm QCD}$ as well as $\mu=5\Lambda_{\rm QCD}$.
The former value is within what is presently being achieved in lattice
calculations for $N_f=2$ \cite{Allton:2003vx}; for $\mu=5\Lambda_{\rm QCD}$
one may expect to be in a quark-gluon phase for all values of the
temperature. In the plots of Fig.~\ref{fig:res15}, the temperature
is varied in both cases between $0.5$ and 5 times $\Lambda_{\rm QCD}$.

The case $\mu=\Lambda_{\rm QCD}$ turns out to be qualitatively
similar to the $\mu=0$ case studied in \cite{Blaizot:2003iq}:
the scale dependence is greatly reduced by keeping $m_E^2$
unexpanded, and the optimization using FAC is close to PMS (where
the latter exists). For $T\lesssim \Lambda_{\rm QCD}$ the scale
dependence increases strongly so that one probably should
conclude that any predictivity is lost. On the other hand,
all optimized results are very close to each other for $T\gtrsim
1.5 \Lambda_{\rm QCD}$ and may be considered as rather definite
predictions.

In the case $\mu=5\Lambda_{\rm QCD}$, the situation is
similar for $T\gtrsim
1.5 \Lambda_{\rm QCD}$, but the scale dependences turn out
to increase more rapidly as $T$ is lowered; moreover,
keeping $m_E^2$ unexpanded does no longer help
to improve this. Yet, the two FAC results are fairly close to each
other for $T\gtrsim 0.5\Lambda_{\rm QCD}$, but show a run-away
behavior for $T$ slightly smaller than $0.5\Lambda_{\rm QCD}$ (not displayed).
This could indicate a beginning breakdown of dimensional reduction,
or at least of its predictivity. On the other hand,
the FAC results appear to remain predictive for $T\gtrsim
1.5 \Lambda_{\rm QCD}$ even when $\mu\sim 5\Lambda_{\rm QCD}$.
At much higher $\mu$, however, $T$ needs to be correspondingly
higher in order that reasonable FAC results are obtained, but this
is primarily due to the run-away behavior of the FAC scales and not
so much due to the behavior of the pressure itself.

\begin{figure}[t]
\centerline{\includegraphics[viewport=50 240 530 540,width=0.9%
\linewidth]{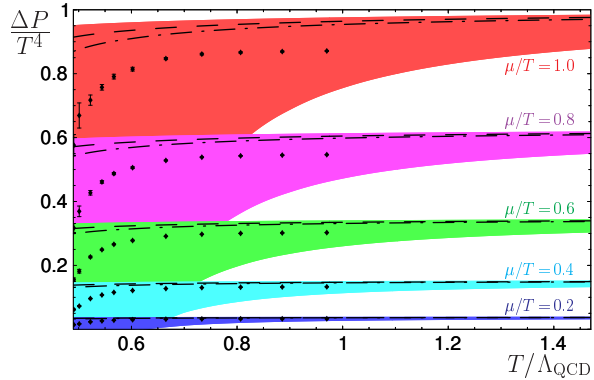}}
\caption{The difference $\Delta P=P(T,\mu)-P(T,0)$ divided by $T^4$ using
the unexpanded three-loop result from dimensional reduction
of Ref.~\cite{Vuorinen:2003fs} for $\mu/T=0.2,\ldots,1.0$
(bottom to top).
Shaded areas correspond to a variation
of $\muMS$ around the FAC-m choice by a factor of 2; dashed and
dash-dotted lines
correspond to the FAC-g and FAC-m results, respectively. 
Also included are the recent lattice data of Ref.~\cite{Allton:2003vx}
(not yet continuum-extrapolated!) assuming $T_c^{\mu=0}=
0.49\, \Lambda_{\rm QCD}$ \cite{Gupta:2000hr}.
\label{fig:dpr}}
\end{figure}

Another popular choice of renormalization scales is the
BLM prescription \cite{Brodsky:1983gc}, which amounts
to simply keeping the FAC scales as obtained in the $N_f\to\infty$ limit.
These scales show no run-away behaviour, but are significantly
smaller than those of FAC-g and FAC-m at finite $N_f$, with
results roughly at the lower boundaries of the bands of
Fig.~\ref{fig:res15}, i.e.~also comparatively far from PMS.

We also note in passing that just as in the $N_f\to\infty$ limit
the two FAC scales happen to be simply proportional
to each other when $N_f=2$: 
$\muMS^{\mbox{\scriptsize FAC-g}}/\muMS^{\mbox{\scriptsize FAC-m}}
=e^{5/29}\approx 1.188$,
while for $N_f\not=2$ this ratio has nontrivial $\mu/T$-dependence.

In Fig.~\ref{fig:dpr} we display the optimized results for the
difference $\Delta P=P(T,\mu)-P(T,0)$ for various
$\mu/T$ corresponding to the recent lattice results
given in Fig.~6 of Ref.~\cite{Allton:2003vx} assuming $T_0\equiv T_c^{\mu=0}=
0.49\, \Lambda_{\rm QCD}$ \cite{Gupta:2000hr}.
At $T/T_0=2$, the highest value considered in \cite{Allton:2003vx},
our FAC-g and FAC-m results exceed the not-yet-continuum-extrapolated 
lattice data consistently by 10.5\% and 9\%, respectively.
This is in fact roughly the expected discretization error
\cite{Karsch:2000ps}.
When normalized to the free value $\Delta P_0$ instead of $T^4$, the results
would be essentially $\mu$-independent and thus also very similar to
the 
$N_f=2+1$ lattice
results of Ref.~\cite{Fodor:2002km} 
as well as the quasiparticle model
results of Refs.~\cite{Szabo:2003kg,Rebhan:2003wn}.

To summarize, we have found that the requirement
of fastest apparent convergence of effective-field-theory parameters
works remarkably well when the perturbative 3-loop result
for the pressure of QCD at finite temperature and chemical
potential is compared with the exactly solvable large-$N_f$
limit, except for the region $T\lesssim 0.1\mu$. 
The perturbative results for finite $N_f$ turn out
to agree reasonably well with existing lattice data
for deconfined QCD with non-zero $\mu$,
and also the otherwise strong renormalization scale dependence
is brought under control, when effective-field-theory parameters
are kept unexpanded as proposed in 
Refs.~\cite{Kajantie:2002wa,Blaizot:2003iq}.


A.~Ipp has
been supported by the Austrian Science Foundation FWF,
project no. 14632-PHY;
A.~Vuorinen has been supported by the Magnus Ehrnrooth Foundation and the Academy
of Finland, contract no. 77744.


\end{document}